\documentclass[%
reprint,
superscriptaddress,
 amsmath,amssymb,
 aps,
 prl,
showkeys]{revtex4-2}

\usepackage{graphicx}% Include figure files
\usepackage[caption=false]{subfig}
\usepackage{dcolumn}% Align table columns on decimal point
\usepackage{bm}% bold math
\usepackage{mathtools}
\usepackage{xcolor,soul}
\usepackage{chngcntr}

\newcommand{\NIST}{
National Institute of Standards and Technology, 325 Broadway, Boulder, Colorado 80305, USA}
\newcommand{\CU}{
Department of Physics, University of Colorado, Boulder, Colorado 80309, USA}

\begin{document}

\preprint{APS/123-QED}

\title{Ratchet Loading and Multi-Ensemble Operation in an Optical Lattice Clock}%

\affiliation{\NIST}
\affiliation{\CU}
\affiliation{Electrical, Computer \& Energy Engineering, University of Colorado, Boulder, Colorado, USA}
\affiliation{National Metrology Institute of Japan (NMIJ), National Institute of Advanced Industrial Science and Technology (AIST), 1-1-1 Umezono, Tsukuba, Ibaraki 305-8563, Japan}
\affiliation{Department of Physics, The Pennsylvania State University, University Park, Pennsylvania 16802, USA}

\author{Y.S. Hassan}
\affiliation{\NIST}
\affiliation{\CU}

\author{T. Kobayashi}
\affiliation{\NIST}
\affiliation{National Metrology Institute of Japan (NMIJ), National Institute of Advanced Industrial Science and Technology (AIST), 1-1-1 Umezono, Tsukuba, Ibaraki 305-8563, Japan}

\author{T. Bothwell}
\affiliation{\NIST}

\author{J.L. Seigel}
\affiliation{\NIST}
\affiliation{\CU}

\author{B.D. Hunt}
\affiliation{\NIST}
\affiliation{\CU}

\author{K. Beloy}
\affiliation{\NIST}

\author{K. Gibble}
\affiliation{\NIST}
\affiliation{Department of Physics, The Pennsylvania State University, University Park, Pennsylvania 16802, USA}

\author{T. Grogan}
\affiliation{\NIST}
\affiliation{\CU}

\author{A.D. Ludlow}
\email{andrew.ludlow@nist.gov}
\affiliation{\NIST}
\affiliation{\CU}
\affiliation{Electrical, Computer \& Energy Engineering, University of Colorado, Boulder, Colorado, USA}

\keywords{Ultra-cold atoms; Magneto-Optical trap; Optical lattice; Atomic clock; Multi-ensemble; Atomic collisions; Ratchet loading}

\date{\today}

\begin{abstract}

We demonstrate programmable control over the spatial distribution of ultra-cold atoms confined in an optical lattice. The control is facilitated through a combination of spatial manipulation of the magneto-optical trap and atomic population shelving to a metastable state. We first employ the technique to load an extended (5 mm) atomic sample with uniform density in an optical lattice clock, reducing atomic interactions and realizing remarkable frequency homogeneity across the atomic cloud. We also prepare multiple spatially separated atomic ensembles and realize multi-ensemble clock operation within the standard one-dimensional (1D) optical lattice clock architecture. Leveraging this technique, we prepare two oppositely spin-polarized ensembles that are independently addressable, offering a platform for implementing spectroscopic protocols for enhanced tracking of local oscillator phase. Finally, we demonstrate a relative fractional frequency instability at one second of $2.4(1) \times10^{-17}$ between two ensembles, useful for characterization of intra-lattice differential systematics. 

\end{abstract}

\maketitle

\section{Introduction}

Optical lattice clocks (OLCs) continue to advance the forefront of frequency metrology owing to their state-of-the-art accuracies \cite{ushijima2015cryogenic,mcgrew2018atomic,bothwell2019jila} and stabilities \cite{schioppo2017ultrastable,oelker2019demonstration}. Their exceptional metrics have enabled novel applications beyond frequency metrology such as geodesy \cite{lisdat2016clock,grotti2018geodesy}, searches for dark matter \cite{wcislo2018new,kennedy2020precision,kobayashi2022search}, and tests of fundamental physics \cite{takamoto2020test,safronova2018search}. Central to OLCs stability and accuracy are carefully confined ensembles of neutral atoms, trapped in `magic' wavelength optical lattices where the differential polarizability between the clock states vanishes. OLC performance is often impacted by (1) quantum projection noise (QPN) \cite{itano1993quantum} or (2) Dick noise \cite{dick1989local}. The QPN limited stability can decrease from efforts to reduce shifts attributed to atomic collisions in the optical lattice by deliberately reducing the loaded atom number \cite{ludlow2015optical}. Dick noise arises from the aliasing of local oscillator (LO) noise by the experimental cycle period, substantially degrading clock stability \cite{dick1989local}. Clocks typically operate with a balance between atom number sufficient to reduce QPN below Dick noise while also seeking to avoid deleterious atomic-interaction-induced frequency shifts.

In parallel with traditional OLC operation, new techniques for addressing LO noise are emerging. The use of two OLCs has allowed a variety of demonstrations for addressing Dick noise limitations such as zero-dead time operation \cite{biedermann2013zero,schioppo2017ultrastable}, quantum non-demolition measurements \cite{bowden2020improving}, and dynamical decoupling \cite{dorscher2020dynamical}. Recent work in 1D OLCs demonstrated a so called quadrature-Ramsey technique, wherein multiple Sr ensembles allowed a proof-of-principle extended Ramsey interrogation \cite{zheng2024reducing}. In related work, Sr atoms in optical tweezers with spatial control allow probing both quadratures of the clock laser phase \cite{shaw2024multi}, also extending atom-light interaction times. A important goal remains: to leverage extended-interrogation techniques within state-of-the-art 1D OLCs without compromising clock accuracy.

\begin{figure*}
\centering
\includegraphics[width=0.98\textwidth]{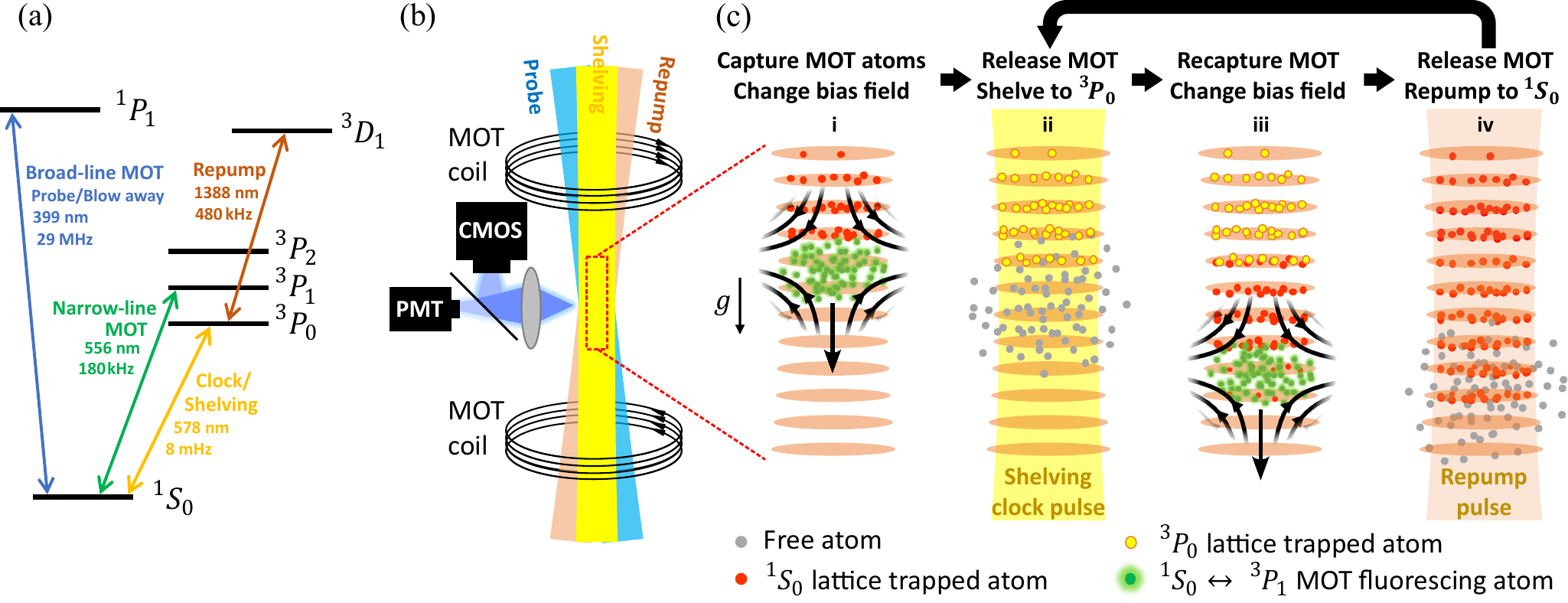}
\caption {(a) Relevant transitions in alkaline earth (-like) atoms, showing wavelengths and linewidths in the case of $\prescript{171}{}{\mathrm{Yb}}$. (b) Experimental layout. The fluorescence detection probe beam and repump are oriented vertically to ensure intensity homogeneity across extended atomic samples. The CMOS camera and a PMT share the fluorescence probe signal through a 50:50 beamsplitter. (c) The ratchet lattice loading scheme: (i) Cold atoms from a standard MOT based on the narrow-line $\prescript{1}{}S_0 \leftrightarrow \prescript{3}{}P_1$ transition (green) are loaded into the lattice (red). The MOT quadrupole zero-field location (at the center of the black magnetic field lines) is programmed along the lattice direction. The cold atomic cloud follows the zero-field location, loading more atoms into new lattice sites. (ii) The MOT is extinguished, and the lattice loaded atoms are shelved to the metastable $^3P_0$ state (yellow). Due to atomic motion, atoms released from the MOT (gray) are only weakly excited to the $^3P_0$ state due to Doppler-broadening. (iii) The narrow-line MOT is turned on to recapture the freely expanding atomic cloud, further cooling and loading them to a new location along the lattice. The shelved atoms are dark to the MOT cycling transition, and are thus immune from heating out of the lattice potential. (iv) The MOT is released, and the shelved atoms are repumped to the ground state to prepare for following lattice load. The repump step is optional if there is no more lattice loading that follows.}
\label{fig:Scheme}
\end{figure*}

In this paper, we demonstrate a new optical lattice loading technique that enables programmable control of the atomic distribution within a 1D OLC that supports state-of-the-art accuracy \cite{mcgrew2018atomic}. The technique allows loading arbitrary atomic cloud distributions along the lattice axis by manipulating the position of the preceding magneto-optical trap (MOT) cloud during the loading process. This control provides immediate benefits: loading extended atomic distributions along the lattice enables operation with high atom number at low density, reducing QPN while mitigating shifts induced by atomic interactions. Moreover, the technique can be utilized to load spatially resolved ensembles of atoms along the lattice, potentially beneficial for emerging techniques for probing beyond the LO coherence time \cite{zheng2024reducing,shaw2024multi}. For example, we demonstrate in-situ synchronous spectroscopy between two ensembles to eliminate the Dick effect by rejecting shared LO noise \cite{bothwell2022resolving,zheng2022differential} as well as many shared systematic perturbations and noise processes. The experimental possibilities of the technique include boosting clock performance by engineering spectroscopic sequences that optimally utilize clock resources \cite{rosenband2013exponential,borregaard2013efficient} or reducing atomic preparation time between measurements \cite{bowden2020improving}. Further, the loading technique is widely applicable to atomic species utilizing a narrowline MOT and metastable state for shelving \cite{petersen2020spectroscopy,patscheider2021observation}.

\section{The Ratchet loading technique}
The technique, which we call `ratchet loading', is based on two main ingredients: (1) control of the narrow-line MOT cloud position with a magnetic bias field and (2) shelving atomic populations to a metastable state that is dark to the MOT cycling transition. The shelving process is analogous to a ratchet as it locks the atoms in their loaded positions, protecting them from scattering MOT photons. Both ingredients are central to many emerging quantum information platforms \cite{chen2022analyzing}, making this technique broadly applicable beyond OLCs. As described later, shelving to a metastable state is not strictly required to produce some of atomic distributions demonstrated in this paper. The experimental apparatus featured here utilizes ultra-cold ytterbium in a 1D optical lattice \cite{mcgrew2018atomic} and has been minimally modified to accommodate the ratchet loading process. First, we installed a CMOS camera that images the atomic cloud perpendicular to the 1D optical lattice. Second, we increase the narrow-line MOT beam diameters ($1/e^2$ intensity)  to 10 mm, ensuring intensity homogeneity across the target atomic cloud control range. Third, we oriented the resonant probe beam used for fluorescence detection of the atoms to be nearly coaxial with the optical lattice, ensuring homogeneous scattering of photons throughout axially-distributed sample. We note that that this system (called here Yb-2) utilizes an optical enhancement cavity for the lattice that can achieve trap depths in excess of 600 $E_r$ ($E_r=\frac{h^2}{2m\lambda^2}$, where $\lambda$ is the lattice wavelength, $m$ is the mass of the $\mathrm{^{171}Yb}$ atom, and $h$ is Planck's constant). The lattice diameter ($1/e^2$ intensity) is 340 $\mu$m.

\begin{figure*}
\centering
\includegraphics[width=0.90\textwidth]{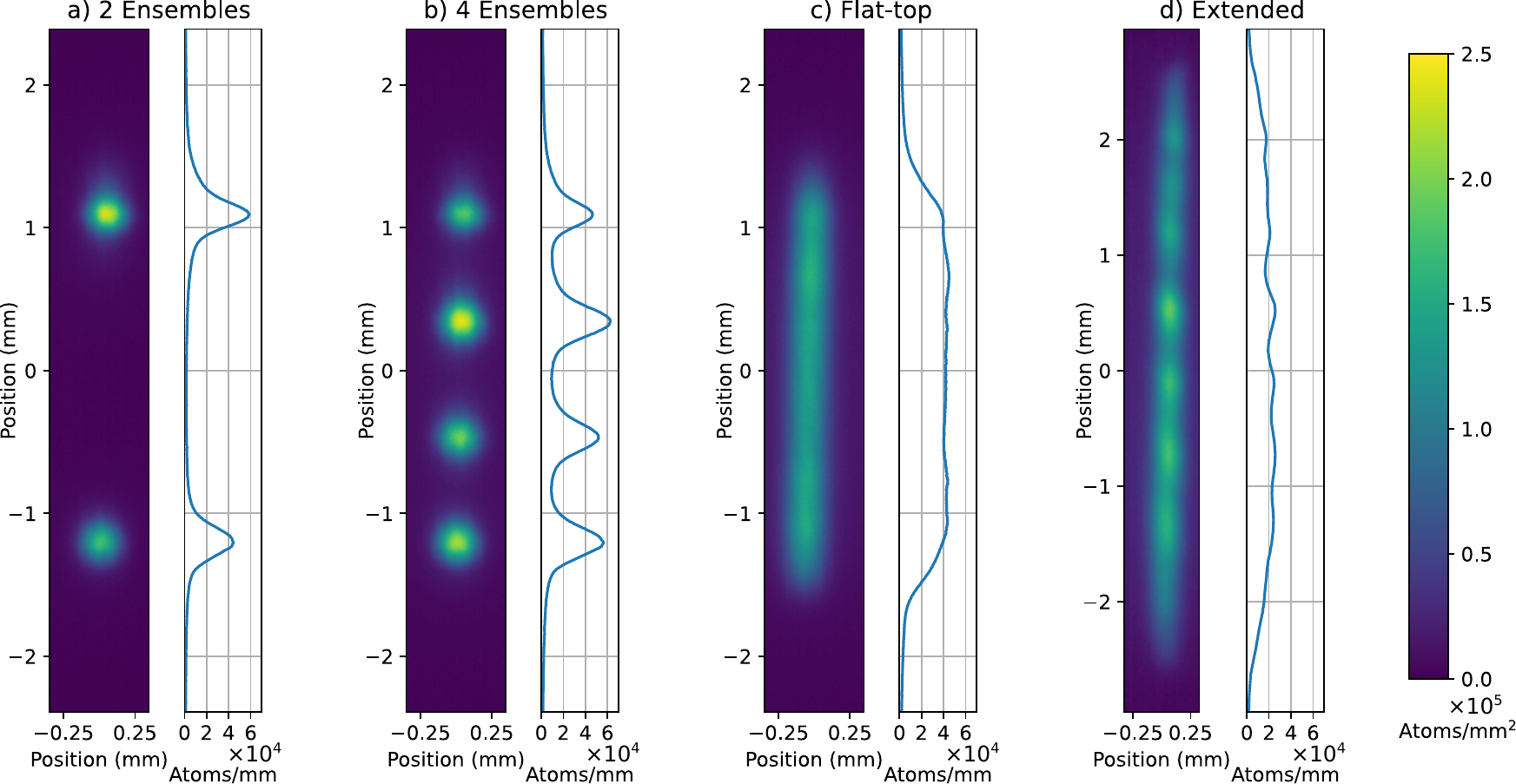}
\caption{\label{fig:Control} Examples of atomic cloud distributions produced using the ratchet loading technique. The left side of each sub-figure shows a 2D density profile, the right side shows the integrated line density profile. a) Two ensembles prepared with different atom number, produced by stepping the MOT bias field between two setpoints.  The top (bottom) ensemble contain $\sim$ 17,000 (12,000) atoms. b) Loading four ensembles by stepping the bias field to four constant setpoints, with different atom number in each ensemble (ranging from 10,000 to 20,000 atoms each). c) Flat-top distribution, providing uniform density over 2.8-mm full width at half maximum, containing $>10^5$ atoms. d) Highly extended distribution, reaching $\sim$ 5 mm and containing in excess of $10^5$ atoms (note the different axes scale). For clarity, the sensitivity of the color scale of the extended distribution is twice that of the other panes. For each distribution, the vertical zero position has been shifted to roughly mark the cloud geometric center.}
\end{figure*}

The ratchet loading scheme, based on positional control of the MOT using magnetic fields, is illustrated in Figure \ref{fig:Scheme}. After capturing atoms in the broad-line $\prescript{1}{}S_0 \leftrightarrow \prescript{1}{}P_1$ MOT, the atoms are further cooled through three stages of narrow-line MOT based on the $\prescript{1}{}S_0 \leftrightarrow \prescript{3}{}P_1$ transition prior to loading into the optical lattice. The third, most closely detuned and lowest intensity stage narrow-line MOT, is the workhorse of the ratchet scheme, referred to as the MOT throughout the remainder of the text. The MOT's detuning, saturation, and magnetic field-zero location determines where and how efficiently the atoms are transferred to the lattice. We optimize these three parameters to load only a subset of atoms, allowing us to load the lattice multiple times. We exploit this control over the MOT to engineer the spatial distribution of the atoms along the lattice. By moving the zero-point of the quadrupole magnetic field to a new lattice location for each loading cycle, we control where atoms are loaded within the optical lattice. Three sets of coils in Helmholtz configuration producing bias fields in three orthogonal directions facilitate controlling the quadrupole zero-field-point. To avoid scattering lattice trapped atoms in a previously loaded location, those atoms are 'shelved' to a dark metastable state---the technique's second ingredient. By shelving prepared lattice atoms into the $\prescript{3}{}P_0$ clock state, they become uncoupled from the MOT cycling transition, preserving their location within the lattice, similar to the shelving function in the stochastically-loaded tweezer system of Ref.~\cite{shaw2023dark}. The $\prescript{3}{}P_0$ atoms can be repumped and shelved as needed, which combined with control over the position of the MOT cloud, enables arbitrary control of the atomic distribution along the lattice.

To shelve atomic population into the metastable $\prescript{3}{}P_0$ state, we utilize an adiabatic rapid passage (ARP) pulse \cite{loy1974observation} to robustly transfer more than 90\% of the lattice trapped ground state population to the metastable state. Due to Doppler effects from atomic motion, atoms not yet trapped into the optical lattice are only weakly excited. The clock ARP pulse functionally resembles a ratchet, locking the trapped atoms in their locations in preparation for the next lattice load, while not affecting the untrapped atoms recently released from the MOT. During the 3.5 ms shelving pulse, the MOT is extinguished to avoid AC Stark shifts or broadening of the excitation spectrum. After the shelving ARP pulse, the MOT beams are turned on to first re-capture the remaining untrapped ground state atoms and then move them to a new location along the lattice, leaving the $\prescript{3}{}P_0$ atoms trapped and undisturbed. The shelving process can be repeated to reload more atoms at additional lattice locations, requiring a repump pulse (followed by radiative decay back to $\prescript{1}{}S_0$) before each subsequent shelving pulse. The repump ensures that the previous metastable state atoms are returned to the ground state before a new iteration of atom shelving (Figure 1).

\section{Programmable spatial atomic distributions}
Variations of the ratchet technique parameters enable programmable control of atomic distributions as shown in Fig.~2. The varied parameters include the bias field's temporal profile, number of shelving and repump pulses, and MOT laser detuning. As an example, we start with a distribution of two spatially resolved atomic ensembles, produced by positioning the MOT cloud to load the top ensemble, briefly releasing the MOT and shelving the atoms to $^3P_0$, then stepping the bias field to recapture the released atoms at the lower position to load the bottom ensemble. This easily implemented level of control demonstrates a 2.3 mm separation between the two ensembles (Fig.~2a), immediately providing spatially separated regions for experiment, with only 30 ms additional preparation time to load the second ensemble. By tuning the shelving pulse efficiency and the transverse overlap of the quadrupole zero-point with the lattice, we can favor loading in one ensemble over the other, thus controlling the relative atom number in each ensemble and providing a potentially convenient platform for evaluating density shifts by relative comparisons between the two ensembles \cite{aeppli2022hamiltonian}. The method can be readily extended to load $N$ atomic ensembles with $N$ bias field steps, using $N-1$ shelving and repump pulses (Figure 2b). Figure \ref{fig:Magfieldprofile} in the appendix details the programming used to produce the atomic distributions shown in Figure \ref{fig:Control}. 

The ratchet technique affords additional control that leads to smearing of the atomic distribution along the lattice instead of loading at localized spots. This is accomplished by ramping the quadrupole-field zero during the MOT loading step instead of holding its position constant between the shelving pulses. With a ramp set at a sufficiently fast rate of $\sim 10\ \mathrm{mm/s}$, a fraction of the lattice trapped atoms do not scatter enough photons to escape the lattice potential and follow the MOT cloud. As the MOT cloud is scanned along the lattice,  continuous loading of the lattice results, producing an atomic smear with its density controlled by a combination of MOT parameters, ramping speed, and lattice trap depth. For example, using a 560 $E_r$ deep lattice, we load a smeared sample with uniform density and vertical length up to 0.8 mm without having to shelve the atoms to the $^3P_0$ state. By stitching 8 of these smeared samples together using 7 shelving and repump pulses, we load long, continuous distributions of more than $10^5$ Yb atoms, shown in Fig.~\ref{fig:Control}c and \ref{fig:Control}d. In Fig.~\ref{fig:Control}c, we target a high but uniform density profile, whereas in Fig.~\ref{fig:Control}d, we aim to produce the maximum vertical extent distribution possible while preserving most of the atoms. In both samples, we adjust the ramp extent and manually tune the bias fields in the transverse directions to control the density profile along the sample by affecting the MOT cloud overlap with the lattice, and thus the loading efficiency of each of the constituent smeared samples. Shallower lattice traps are less capable of holding the loaded atoms during the MOT field ramp, and thus produce shorter smeared samples compared to deeper traps. However, the same distributions can be achieved by stitching more of the shorter smeared samples and shelving pulses. 

As a simpler variation of the standard ratchet loading protocol in the case of spatially separate ensembles, we found that if the vertical bias field is stepped far enough after loading one ensemble, there is no need to shelve the atoms to the $\prescript{3}{}P_0$ state to load the following ensemble. This effect arises solely due to the large local field of a few hundred $\mu$T at the loaded ensemble location, which induces a Zeeman shift large enough to cause the loaded ensemble to become dark to the MOT cycling transition. This extension of the ratchet loading technique broadens the applicability to species that lack access to metastable states. In our setup, the minimum value of the step corresponds to $\sim $ 3 mm separation between the two ensembles, providing an effective detuning due to the Zeeman shift of $ 20 \times \Gamma' \sim $ 4 MHz, where $\Gamma'=\Gamma\sqrt{1+S}$ with natural linewidth $\Gamma=$ 180 kHz and $S\sim 1$ is the saturation parameter of the MOT.

\begin{figure}[b]
\centering
\includegraphics[width=1.0\columnwidth]{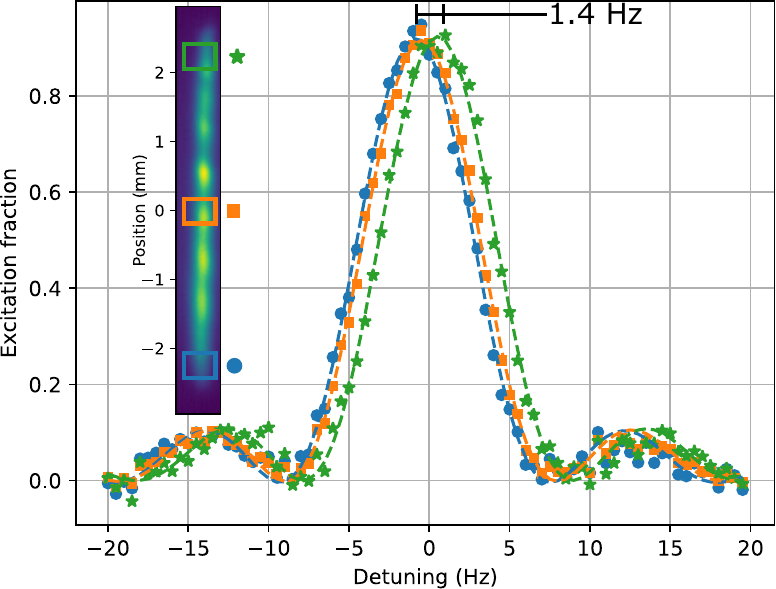}
\caption{\label{fig:ClockStuff} Spectral mapping of magnetic field across the highly extended sample. Excitation profile of selected regions using a 100-ms Rabi pulse demonstrating the near uniformity of Zeeman shifts over the sample. The regions show a frequency gradient of $\sim$ 1.4 Hz over 5 mm, corresponding to 7 mG magnetic field difference between the extreme ends of the sample. The dashed curves are Rabi lineshape fits to the data points. The inset shows the selected regions with corresponding colors at the extreme ends and the center of the sample. }
\end{figure}

\section{Lattice clocks: Spectral homogeneity and Density Shifts}
We now demonstrate the utility of ratchet loading for quantum metrology, by implementing the technique in a state-of-the-art optical lattice clock experiment. Extended atomic distributions offer a simple solution for reducing atomic density shifts without compromising the atom number, thereby suppressing QPN \cite{PhysRevLett.132.133201}. For the 5 mm sample demonstrated here, we estimate a mean fractional density shift of $2\times 10^{-19}$ for $10^5$ spin-polarized and 3D-cooled $^{171}\mathrm{Yb}$ atoms at an operational trap depth of 20 $E_r$\cite{mcgrew2018atomic,zhang2022subrecoil}, corresponding to a QPN-limited performance of $<2\times 10^{-18}/\sqrt{\tau}$ for 560 ms Ramsey free evolution time, where $\tau$ is the averaging time in seconds. Alternatively, flat-top distributions offer the advantage of nearly constant density shift for all atoms, offering the prospect of preserving global contrast during clock spectroscopy in the regime of strong interactions. In Fig.~2c, we load a $\sim$ 2.7~mm-flat-top profile that exhibits less than 5\% variation of atomic density for more than 85\% of the loaded atoms. This translates to $<1\times10^{-21}$ fractional density shift variation at the same clock operating conditions. For the large vertical extents demonstrated here, fractional frequency differences at $5\times 10^{-19}$ is expected due to the gravitational red-shift between the top and the bottom of the sample. Such a shift can in principle be corrected for at the  $< 1\times 10^{-19}$ level, given the spatial profile of the sample from the camera \cite{bothwell2022resolving}.

For extended samples, extra care may be needed to ensure magnetic field uniformity along the atomic sample for subsequent quantum control or precision spectroscopy. For example, gradients from nearby magnetic materials (such as the vacuum chamber) can introduce magnetic field inhomogeneities, which might degrade clock stability through reduction of excitation contrast. During clock operation, we apply an 0.1 mT field and cancel the first-order Zeeman shifts by averaging the frequency between the two spin states with spin projections $F=\frac{1}{2}$. The local magnetic field for each ensemble is determined by the frequency difference between the two stretched states with total angular momentum projection $m_F = \pm \frac{1}{2}$, with further corrections applied for second order Zeeman shifts based on the spectroscopically determined local field.  We evaluated stray magnetic field components along the atomic sample. Armed with this knowledge, we installed additional magnetic bias controls, enabling the cancellation of stray gradients in three orthogonal directions. We spectroscopically evaluate the local magnetic field homogeneity to be better than 0.7~$\mu$T across the extremes of the 5 mm atomic sample (Fig.~\ref{fig:ClockStuff}). Clock spectroscopy of the extended sample reveals a global 90\% contrast for a 150 ms long Rabi pulse, essentially undegraded when compared to localized atomic samples.

\begin{figure}[b]
\includegraphics[width=1.0\columnwidth]{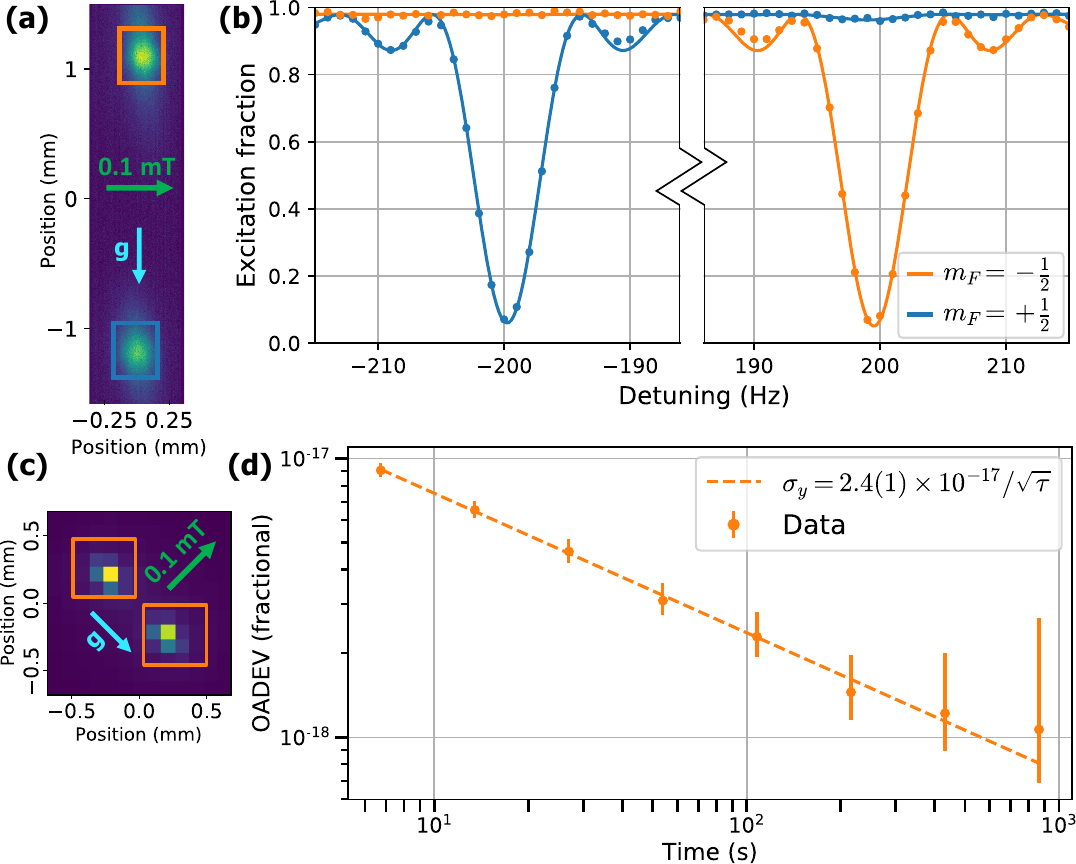}
\caption{\label{fig:spinresults} (a) The spatial profile of two oppositely spin-polarized ensembles in Yb-2. The top (bottom) ensemble is prepared with $m_F=-\frac{1}{2}(+\frac{1}{2})$. The ensembles appear with additional vertical spread due to longer probe exposure compared to the profile in figure \ref{fig:Control}. (b) Inverted narrow-line spectroscopy of the regions highlighted in panel (a) to assess the spectroscopic cross-talk between the two ensembles. By scanning over the resonance of each ensemble, we observe the coherent de-excitation of more than 90\% each and spin purity in excess of 98\%. The measured spin impurity is dominated by fluorescence bleeding between the two integrated spatial regions. Error bars of all points are $<$ 1\%, too small to be displayed on the graph. (c) Spatial profile in Yb-1. The 45\textdegree \ tilt is due to camera orientation. Each ensemble contains $\sim$9,000 atoms. (d) Overlapping Allan deviation of the frequency difference between the two identical ensembles in panel (c). The dashed line is the white frequency noise fit with $\sigma_y=2.4(1)\times10^{-17}/\sqrt{\tau}$}
\end{figure}

\section{Lattice Clocks: Multi-ensemble operation}
As discussed in the introduction, operating a single lattice clock with multiple spatially separated atomic ensembles offers a versatile platform to efficiently allocate atomic resources for the improvement of clock performance \cite{zheng2024reducing}. The independent control and spatially resolved detection of each ensemble enables experimental possibilities that are inaccessible to single-ensemble clocks while preserving the accuracy control of traditional 1D OLCs. Towards this, we first load two spin-polarized ensembles with opposite spins separated by 2.3 mm (preparation details are available in Appendix~A). By applying a constant bias field of 0.1 mT, the $\prescript{1}{}{S_0},m_F=\pm\frac{1}{2} \rightarrow \prescript{3}{}{P_0},m_F=\pm \frac{1}{2}$ $\pi$-transitions are split by 400 Hz, enabling independent ensemble addressing via the sub-Hz linewidth clock light. To assess the spin purity of each sample, we perform narrow-line spectroscopy with a 150 ms Rabi pulse on a sideband-cooled sample in a 120 $E_r$ deep trap and utilize ensemble-resolved imaging to measure the excitation of each ensemble as a function of laser frequency. We observe spin purity in excess of 98\% for each ensemble, detection limited by fluorescence bleeding between the two integrated regions (see Fig. \ref{fig:spinresults}). We anticipate that  a better imaging system would allow a better bound on the spin purity \cite{kaufman2021quantum}.

To demonstrate application of the ratchet loading technique for laser-noise-free frequency comparisons, we employ the technique in a second apparatus (Yb-1) that has an upgraded imaging system. Yb-1, described in Ref.~\cite{mcgrew2018atomic}, employs a retro-reflected lattice with a tighter trap waist and lower trap depths. The system has been upgraded with an electron-multiplying CCD camera, aligned orthogonal to the lattice axis to support in-situ frequency measurements. In Yb-1, we apply the ratchet loading protocol to prepare two spatially separated ensembles in a 60 $E_r$ lattice with a few thousand atoms each. The two ensembles are prepared in identical spin states and cooled to an average axial motional state $\Bar{n_z} \approx 0$ and radial temperature $\approx 1.3$ $\mu$K \cite{ChenClock}. Standard clock operation proceeds, with the LO locked to the full atomic sample via standard global signal detection on a photomultiplier tube \cite{mcgrew2018atomic}. During state readout the camera records the fluorescence signals for region-resolved state detection.

The record of ensemble excitation fractions from the camera combined with the 560 ms Rabi line shape and known contrast give a conversion from excitation fraction to frequency for each ensemble \cite{marti2018imaging,bothwell2022resolving}. The clock frequency of each region is thus differentially compared with rejection of LO noise, probe noise, and other systematics, allowing rapid evaluation of frequency differences and a valuable probe of clock frequency gradients. As shown in Figure 4, we find the fractional frequency difference between regions that averages as $2.4(1)\times 10^{-17}/\sqrt{\tau}$. Extrapolating to half the total time of the 30-minutes-long dataset shown in Fig. \ref{fig:spinresults}d, we find a fractional frequency difference of $4(8)\times10^{-19}$, consistent with zero, after correcting only for the first-order Zeeman shift. The utilization of these two samples for evaluation of lattice light shifts \cite{brown2017hyperpolarizability,ushijima2018operational,nemitz2019modeling,kim2023evaluation}, is currently under study.

\section{Conclusion}
We demonstrate a novel lattice loading technique for OLCs that enables engineered atomic cloud distributions within the lattice. Our technique requires minimal experimental modification and is readily employed in traditional OLCs as well as many other atomic systems. We demonstrate loading highly extended atomic samples and multiple spatially resolved atomic ensembles. The former offers a solution for loading large atom numbers while maintaining low atomic density, facilitating future clocks achieving QPN as they continue to advance. The latter paves the way for more efficient use of atomic resources by enabling experimental platforms not accessible to single-ensemble clocks. Although we use 1D-optical lattices for our demonstrations, we anticipate that the technique can be extended to 2D and 3D optical lattices of varying geometries. Using a distribution with two spatially resolved ensembles, we demonstrate independent addressability of two spin-polarized ensembles exhibiting $<$ 2\% cross-talk as well as relative frequency stability of $2.4(1)\times 10^{-17}$ at 1 s of averaging time. Both addressability and high relative stability offer a platform for evaluating some clock systematics (e.g., lattice light shifts) through differential comparisons that reject LO noise and various systematic effects. Finally, the ratchet loading platform lends itself to new techniques that call for engineered interrogation sequences on multiple clock ensembles to extend atomic interrogation beyond LO coherence time, all while maintaining compatibility with accuracy-focused 1D OLC operation.

\section{Acknowledgments}
We gratefully acknowledge C.-C. Chen and W. F. McGrew for useful discussions on the experimental implementation. We also thank J. A. Sherman and W. J. Dworschack for careful reading of this manuscript. This work was supported by NIST, ONR, and NSF QLCI Grant No. 2016244 and Grant No. 2012117 (KG). The authors declare no competing interests.

The data that support the findings of this study are available from the corresponding author upon reasonable request.

% \bibliography{references}
% \bibliographystyle{unsrt}

\appendix

\counterwithin{figure}{section}
\renewcommand{\thefigure}{A\arabic{figure}}

\section{Appendix A: Preparation of the spin-polarized sample}
To load oppositely spin-polarized samples, we start with two unpolarized ensembles in the ground state (Fig.~\ref{fig:spinprepNmot}a). We apply a large magnetic field gradient ($\sim$10 G/cm) in the vertical direction so that each ensemble experiences a different local magnetic field.  We adjust the bias field in all three directions to achieve a local field that points in the downward vertical direction, but with a different magnitude for each ensemble. We then use the clock light to selectively excite one spin state for each ensemble. The selection is performed through a 3.5 ms adiabatic rapid passage (ARP) pulse ($\Omega_R/2\pi\approx$ 3 kHz) that sweeps by $\sim$800 Hz over a single spin feature for each ensemble, selectively exciting the $m_F=-\frac{1} {2}$ state for the upper ensemble and the $m_F=+\frac{1}{2}$ state for the lower ensemble, as in Fig.~\ref{fig:spinprepNmot}b. We then clear the remaining ground state atoms through a resonant 399-nm pulse and perform narrow-line inverted spectroscopy to assess the spin purity of the prepared ensembles. We opted to use the narrow clock light for spin selection to ensure spin purity of the prepared ensembles. It is also possible to optically pump the spin states on the $\prescript{1}{}S_0 \leftrightarrow \prescript{3}{}P_1$ transition, but extra care must be taken to independently address each ensemble owing to the broader line-width of the transition.

\section{Appendix B: Evaluating stray magnetic field gradient components}
To spatially characterize the stray magnetic field components along the sample, we apply a large external magnetic field along a particular direction. The sample atoms experience first order Zeeman shift due to the total magnetic field, given by the vector sum of the stray field and the applied field. The applied external field is typically $\sim$ 100 times larger than the stray field. Thus the stray field component parallel to the applied field contributes meaningfully to the Zeeman shift, while the perpendicular components are suppressed in quadrature. The application of the large external field effectively isolates the stray field component in its direction. Using spatially resolved spectroscopy, we can bound the gradients of the stray field component across the sample. We repeat the measurement for all three Cartesian directions and observe the vertical component of the field to exhibit the largest change along the sample. We installed a single coil on top of the chamber to cancel the vertical-component gradient. The uncorrected field gradients in the other two directions dominate the frequency gradient shown in Fig. \ref{fig:ClockStuff}.

\section{Appendix C: Second order Zeeman shift}
Even with good contrast and proper cancellation of the first order Zeeman shift via averaging of opposite $m_F$ components and second order Zeeman shift through the spectroscopically determined average magnetic field, inhomogeneous magnetic fields over extended atomic samples produce an error in the evaluation of the second order Zeeman shift of the clock frequency. To estimate the error, we simulate the lock feedback loop between the atoms and the LO for idealized cloud distributions that span the demonstrated distributions, subject to the experimental bias field plus the residual field inhomogeneities observed. We bound the biases due to second order Zeeman shift below $10^{-19}$ fractional frequency from the unperturbed center frequency for all the atomic distributions we explored, which is compatible with state-of-the-art optical lattice clock operation at the low-$10^{-18}$ level.

\section{Appendix D: Shot-to-shot stability of the ratchet loading technique}
We also investigate the shot-to-shot stability of the flat-top profile. By integrating the fluorescence from three segments of the distribution at the top, middle, and bottom of the sample, and normalizing to the total fluorescence, we find that the shot-to-shot relative atom number fluctuations between segments is better than 1.3\% for all segments, better than the total atom shot-to-shot stability of 2.0\%.

\section{Appendix E: Estimation of density shift for the extended sample}
The density shift is estimated based on the measured value for Yb-2 in Ref. \cite{mcgrew2018atomic}, scaled to account for the different atom number and trap conditions highlighted in the text, specifically lattice depth and temperature in both the axial and transverse directions. We define $V$ to be the mean volume occupied by a lattice trapped atom given by:
\begin{equation}
    V=\prod_{j=x,y,z}\sqrt{\frac{\hbar}{m\Omega_j}}\sqrt{2\langle n_j \rangle +1}\ ,
\end{equation}
where $\Omega_j$ is the trap frequency and $\langle n_j \rangle$ is the mean motional quantum occupation number for the $j$-th direction. Given a Boltzmann distribution, $\langle n_j \rangle = (\mathrm{exp}(\hbar\Omega_j/k_B T_j)-1)^{-1}$.
Given the trap conditions in Ref. \cite{mcgrew2018atomic} and those proposed in the main text, we find that $V$ is reduced by $\sim12\times$ compared to Ref. \cite{mcgrew2018atomic}. But for the 5 mm extended sample, the atoms occupy $\sim15\times$ more lattice sites. Further, the mean collision energy of sample is reduced by $\sim7\times$ due to the extra cooling in the transverse dimensions, proportionally suppressing the $p$-wave contribution to density shift due to the centrifugal barrier \cite{ludlow2008strontium}.

\setcounter{figure}{0}

\begin{figure*}
\centering
\includegraphics[width=0.85\textwidth]{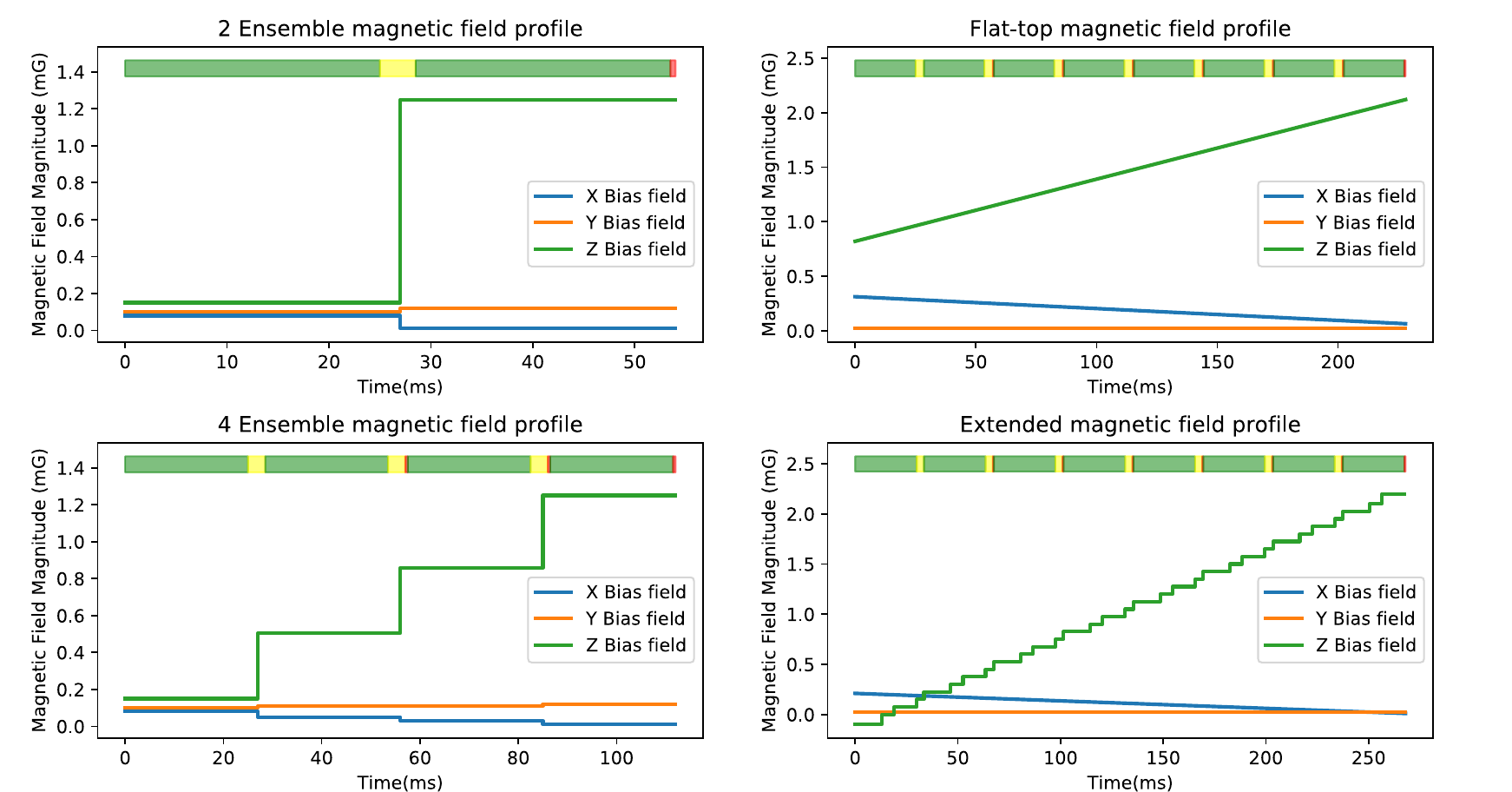}
\caption{\label{fig:Magfieldprofile} Temporal profiles of the bias field programming that is used to produce the atomic spatial profiles in figure 2 of the main text. Given the orientation of the vertical bias field coils, higher field magnitude moves the quadrupole zero-field position downwards. Bars on top of each sub-figure show the time duration of ratchet loading steps: the MOT is active (green), ARP pulse after releasing the MOT (yellow) and repump pulse (red, magnified by 10$\times$ for clarity). The $X$ and $Y$ bias fields need to be stepped/ramped to optimally overlap the MOT zero-field position with the lattice during the $Z$ bias field stepping/ramping. We expect this is due to the lattice orientation being tilted to the MOT quadrupole field vertical axis. The step-like profile of $Z$ bias field for the extended profile best represents a ramp, but is step-like due to technical limitations of the current driver and is not a critical feature of the extended profile.}
\end{figure*}

\begin{figure*}
\centering
\includegraphics[width=1.0\textwidth]{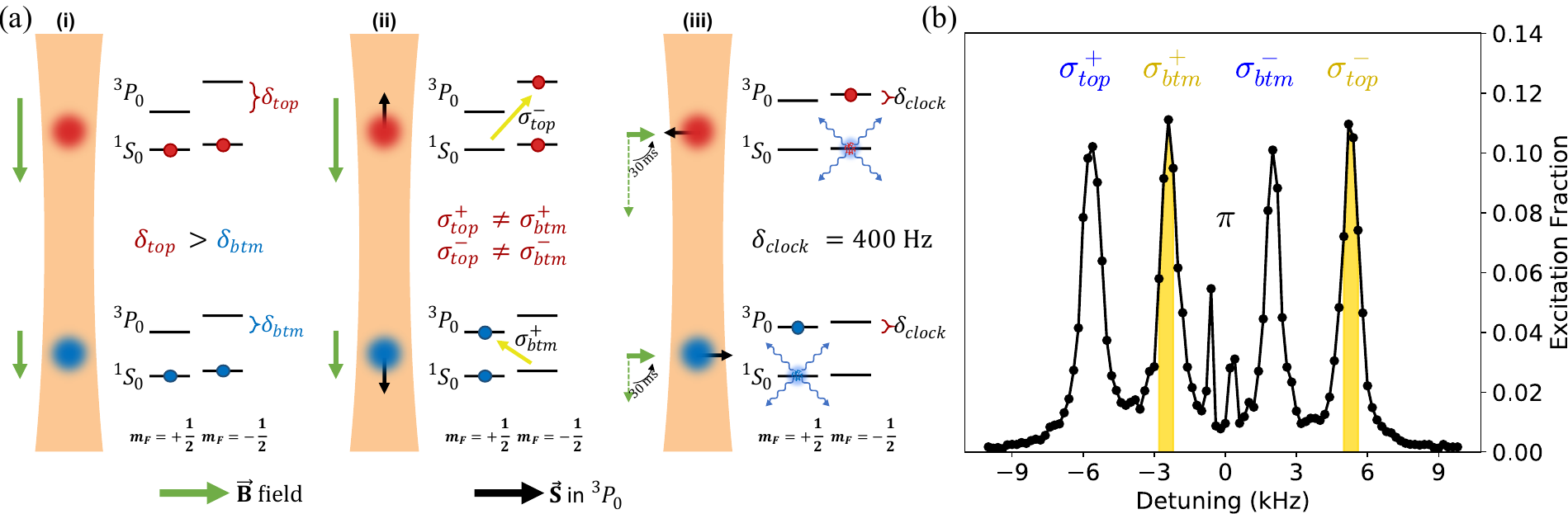}
\caption{\label{fig:spinprepNmot} (a) Spin-polarization scheme for 2-ensemble distribution. (i) MOT quadrupole field of an anti-Helmholtz coil arrangement produces different local magnetic fields for each ensemble. The local field for each ensemble points vertically downwards and is larger for the top ensemble. The bias coils are adjusted to ensure that the field direction is along the lattice, suppressing clock $\pi$ transitions. (ii) Two sequential ARP pulses sweep over the $\sigma^- (\sigma^+)$ transition for the top (bottom) sample, selectively exciting the $m_F=+\frac{1}{2}(-\frac{1}{2})$ state. Due to the different local field magnitudes, each ARP pulse addresses only one ensemble at a time (iii) To rotate the quantization axis while preserving spin purity, the MOT quadrupole field is adiabatically switched to a 0.1-mT field produced by a Helmholtz coil arrangement perpendicular to the lattice axis for spectroscopy. The remaining ground state atoms are then cleared using a resonant 399-nm pulse, leaving pure spin polarized samples in the excited state for inverted clock spectroscopy. (b) Spectrum of two spin-unpolarized ensembles in a quadrupole magnetic field. Due to the larger local magnetic field at the top ensemble ($\sim$ 0.5 mT), its Zeeman splitting is larger compared to the bottom ensemble. The clock $\pi$ transitions are suppressed since the wave vector of the laser field is nearly parallel to the magnetic field direction. The excitation fraction is calculated through fluorescence collected on the PMT, while spatial peak assignment is confirmed by the camera. The yellow shading indicates the frequency sweep of the two ARP pulses that select opposite spin states for each ensemble. The population of the other two peaks are cleared from the lattice by resonant 399-nm pulses.}
\end{figure*}

\end{document}